\documentclass[10pt,journal,final]{IEEEtran}
\usepackage{lineno,hyperref}
\usepackage{balance}
\usepackage{graphicx}
\usepackage{subfigure}
\usepackage{epstopdf}
\usepackage{mathrsfs}
\usepackage{graphicx}
\usepackage{amssymb}
\usepackage{amsmath}
\usepackage{amsfonts}
\usepackage{amsthm}
\usepackage{color}
\usepackage{balance}
\usepackage{algorithm}
\usepackage{algorithmic}
\usepackage{amsmath,amssymb,amsfonts}
\usepackage{algorithmic}
\usepackage{textcomp}
\usepackage{stfloats}
\usepackage[numbers,sort&compress]{natbib}
\usepackage{balance}

\allowdisplaybreaks[4]

\newtheorem{remark}{Remark}
\newtheorem{definition}{Definition}
\newtheorem{assumption}{Assumption}
\newtheorem{lemma}{Lemma}
\newtheorem{proposition}{Proposition}
\newtheorem{theorem}{Theorem}
\newtheorem{corollary}{Corollary}

\newcommand{\norm}[1]{\left\Vert#1\right\Vert}

\newcommand{\Real}{\mathbb R}
\newcommand{\Tran}{\mathrm T}

\begin{document}

\title{Robust Adaptive Safety-Critical Control for Unknown Systems with Finite-Time Element-Wise Parameter Estimation}
	\author{Shengbo Wang, Bo Lyu, Shiping Wen,~\IEEEmembership{Senior Member,~IEEE}, Kaibo Shi, \\ Song Zhu, and Tingwen Huang,~\IEEEmembership{Fellow,~IEEE} 	
	
	\thanks{This publication was made possible by NPRP grant: NPRP 9-466-1-103 from Qatar National Research Fund. The statements made herein are solely the responsibility of the authors. \textit{(Corresponding authors: Shiping Wen.)}
}
	
		\thanks{S. Wang and B. Lyu are with School of Computer Science and Engineering, University of Electronic Science and Technology of China, Chengdu 611731, China (e-mail:  shnbo.wang@foxmail.com; blyucs@outlook.com). S. Wen is with Australian AI Institute, Faculty of Engineering and Information Technology, University of Technology Sydney, NSW 2007, Australia (shiping.wen@uts.edu.au).   K. Shi is with School of Information Science and Engineering, Chengdu University, Chengdu, 611040, China, (email: skbs111@163.com). S. Zhu is with School of Mathematics, China University of Mining and Technology, Xuzhou 221116, China (e-mail: songzhu@cumt.edu.cn). 
		T. Huang is with Science Program, Texas A \& M University at Qatar, Doha 23874, Qatar (e-mail: tingwen.huang@qatar.tamu.edu).}

}

	\markboth{}
{}
	\maketitle

\begin{abstract}
	Safety is always one of the most critical principles for a system to be controlled. This paper investigates a safety-critical control scheme for unknown structured systems by using the control barrier function (CBF) method. Benefited from the dynamic regressor extension and mixing (DREM), an extended element-wise parameter identification law is utilized to dismiss the uncertainty. On the one hand, it is shown that the proposed control scheme can always guarantee the safety in the identification process with noised signal injection excitation, which was not considered in the previous study. On the other hand, the element-wise estimation process in DREM can minimize conservatism of the safe adaptive process compared to other existing adaptive CBF algorithms. The stability as well as the forward invariance of the presented safe control-estimation scheme is proved. Furthermore, the robustness of the scheme under bounded disturbances is analyzed, where a robust CBF with modest conditions is used to ensure safety. The framework is illustrated by simulations on adaptive cruise control, where the slope resistance of the following vehicle is robustly estimated in finite time against small disturbances and the potential crash risk is avoided by the proposed safe control scheme.
\end{abstract}



\section{Introduction}\label{Section_Intro}
As likely the most important principle, safety should be the first consideration in design of a system. With the description of safety through perception or interaction to the environment, the decision making of an autonomous agent attracts the interests of many researches for decades. For instance, the control performance under safe or other constraints is considered by MPC or constrained optimal control method \cite{ANDERSON2019364}, the efficient online controller with stable and feasible real-time response have been analyzed in \cite{tac_mpc_efficient,safeandfast_tsmcs}, the application of the safety-critical control in e.g. robotics can be found in \cite{CBF_overview,tsmcs_CL_ROBOT}. The safety-critical control for unknown systems is rather intractable due the impact of the unknown part of the dynamics. This paper will present a robust control-estimation method for unknown systems with robust and adaptive safety against the uncertainty and disturbances.

\par The control performance of autonomous systems is directly determined by the knowledge of system model. In terms of the structured systems with parameter uncertainty, there are generally two ways to optimize the control performance. One way is to eliminate dependence on the unknown dynamics. Using integral-based reinforcement learning for partially unknown nonlinear system has been proven to be effective and such method can handle input constraints \cite{integral_reinforce_constraint}, discrete control scheme \cite{shnbo_PETC_IRL}, etc. The idea of integral-based constraints has also been used in safety-critical control \cite{intergral_CBF}. Adaptive control e.g. model reference adaptive control works with signal excitation or intelligent data storage to achieve the control objective \cite{IE_MRAC_2013}. Nevertheless, these methods need to be designed separately for different control objectives, and the uncertainty in dynamics of systems always exists. What is worse, the system state is unpredictable owing to uncertainty, which can lead to algorithm conservativeness or great potential safety hazard. The other way is to estimate the unknown parameters using sampled collective data. Through filters or estimation, the regression model of parameterized uncertainty can be obtained and the unknown parameters can be identified under designed criteria, see \cite{Annual_Reviews} for detailed discussions. When the unknown parameters are time-invariant, persistent excitation condition (PE) required for convergence \cite{PEcondition} can be relaxed to interval excitation (IE) one \cite{IEcondition,IE_MRAC_2013,initialexcitation_comparetorank,tsmcs_CL_ROBOT}. Alternatively, important data set and storage can be used to accelerate the approximation process \cite{Rushikesh_tac_concurrent_pestimation, RL_concurrent_tracking_converge,LQRHtracking_unknownnonlinear_Jiang, TCYB_tracking_NN}. The comparison of IE condition and storage based method can be found in \cite{initialexcitation_comparetorank}. A new identification method named dynamic regressor extension and mixing (DREM) is more powerful since it not only relaxes the PE condition without affect the stability, but transforms the vector or matrix estimation task into element-wise scalar estimation task as well \cite{DREM_begin}. The estimation of unknown parameters in matrix form using DREM was recently reported in \cite{matrix_estimation_finite_time} with finite-time convergence, same method was improved to track time-varying parameters in \cite{DREM_improve_time_varying}. A robust DREM with high-gain injected adaptation law was proposed in \cite{finite_time_robust} to achieve finite-time convergence as well as short-time input-to-state stability (ISS) considering disturbances. As an advanced and powerful method for identification, the finite-time DREM is also utilized in this paper to estimate the unknown parameters in structured control systems. 

\par A critical issue is, the parameter identification (typically in an on-line version) needs the regressor signals to be excited, cf \cite{Annual_Reviews,PEcondition,IEcondition,IE_MRAC_2013,initialexcitation_comparetorank,DREM_begin,matrix_estimation_finite_time,DREM_improve_time_varying,finite_time_robust}, and such requirement can lead system state to an unsafe region since the system model is uncertain. For instance, an autonomous cruise vehicle (as discussed in simulation part) is not allowed to collide with the front vehicle during estimation process. Therefore, while introducing the (finite-time) DREM for parameter identification, safety of the systems should be considered at the same time. To ensure the safety with unknown dynamics, an active learning method has been proposed to estimate the reachable sets in \cite{Reachable_TCYB}. However, the computation of reachability takes high complexity and may cause the curse of dimensionality. One of the most convenient and effective methods to ensure safety for autonomous systems is to construct a control barrier function (CBF) as the safety constraint and then optimize the constrained control objective, see \cite{CBF_overview} for an overview of the development of CBF. The CBF-based safe control can be solved in a point-wise manner using quadratic programming (QP). A typically popular result in recent research is the combination of CBF and control Lyapunov function (CLF), which can be applied to real-time control scenario \cite{ACC_example,cbfdefine}. An extension of this method has recently been proposed in \cite{CLFCBF_extended} to get rid of the undesirable equilibrium points with non-smooth CBFs. The robustness of CBF-CLF methods was studied in \cite{robust_cbf} for bounded, known or unknown disturbances. Same method is utilized in this paper to enhance the robustness towards bounded unknown disturbances of the proposed algorithms. In addition, other types of uncertainty with CBF-based safety guaranteed methods can be found in \cite{aCBFCLF_estimation_distur,Unstructured,stochastic_CBF}. For structured uncertainty, estimation was introduced in \cite{aCBFCLF_estimation_distur} to formulate a robust QP. The unstructured uncertainty of second order nonlinear systems was solved in \cite{Unstructured} successfully using CBF-QP. For unbounded, stochastic disturbances, a safety verification method was presented in \cite{stochastic_CBF}. The smooth robust CBF construction is considered in this paper and a QP-based algorithm is used to get the numerical solutions for control input.

\par The concept of adaptive CBF (aCBF) was first presented in \cite{aCBF_first_parameter}, in which parameter uncertainty was considered and dismissed by CBF-based adaptation law. Same result was extended to reduce the conservativeness using a tightened aCBF and a data-driven approach in \cite{RaCBF_IEEECSL}. To guarantee feasibility of this method with time-varying control bounds and noise, a relaxation-based aCBF was studied in \cite{aCBF_timevarying_noise} for high relative degree system dynamics. Unfortunately, for parameterized uncertainty, these methods can not guarantee the accurate or robust estimation for unknown parameters. To learning the system dynamics while ensuring safety, CBF together with Bayesian Learning methods were introduced for second order dynamics in \cite{acbf_Bayesian} and for high order systems in \cite{acbf_highorder_beisienlearning}. The safe parameter estimation process was considered in \cite{safe_uncertain_ELsystem} for uncertain Euler-Lagrange systems, where the parameter estimation law converged slowly and was less robust to disturbances. A fixed-time adaptation law with aCBF has been studied in \cite{fixedtimeACBF}, where a modified version of aCBF was utilized to describe the estimation process and reduce the conservativeness of the tightened aCBF introduced in \cite{RaCBF_IEEECSL} during adaptation. However, the robustness of the complex filter-based learning algorithm in \cite{fixedtimeACBF} is lack of research. In contrast, the (finite-time) DREM with element-wise estimation can be more appropriate for safe and robust adaptation with precise parameter identification.

\par Based on these observations, this paper investigates the robust safety-critical control and estimation framework for structured unknown systems by combining finite-time DREM and robost CBF-QP method. The main contributions are listed below.
\begin{enumerate}[\IEEEsetlabelwidth{12)}]
	\item The finite-time parameter estimation for structured control affine systems is introduced through robust DREM which was presented in \cite{finite_time_robust}. Differently, the unknown parameters are in a matrix form other than a vector one. Compare to other learning or regression algorithms \cite{Annual_Reviews,DREM_begin,initialexcitation_comparetorank,shnbo_PETC_IRL,integral_reinforce_constraint,matrix_estimation_finite_time,DREM_improve_time_varying,finite_time_robust}, the noise signals generated to ensure PE or IE condition is filtered by aCBF method to keep control system safe during the estimation process.
	\item A modified aCBF method is proposed to ensure safety of the unknown systems. With the element-wise parameter estimation by DREM, the worst-case estimation error bound of each unknown parameters can be computed and converge to zero in a finite time, which will eliminate the conservative of aCBF due to uncertainty. Analysis of stability and forward invariance of the proposed algorithms are presented as well.
	\item It is founded that, during the estimation process introducing the switched aCBF constraints can minimize the conservative of aCBF methods compared to \cite{RaCBF_IEEECSL,fixedtimeACBF}. Furthermore, the robustness of the presented methods is analyzed theoretically. In simulations, a practical problem of adaptive cruise control (ACC) is given to illustrate the effectiveness of the proposed algorithms, where the slope resistance is estimated robustly without collision accident.
\end{enumerate}
	
\par \emph{Notation}: For a matrix $A\in \Real^{n\times n}$,  $det\{A\}$, $tr(A)$, $adj\{A\}$, $A^\mathrm{T}$ and $A^{-1}$ represent the determinant, trace, adjoint matrix, transpose matrix and the inverse matrix of $A$, respectively.
For a signal $\delta(t)\in \Real$, denote $\delta(t) \in \mathcal{L}_2$ if $\int_{-\infty}^{+\infty} \delta^2(t)\, dt < +\infty$. A continuous function $\alpha: \Real_+\to \Real_+$ is a class $\mathcal{K}_{\infty}$ function if $\alpha(0) = 0$ and strictly increasing to infinity. A continuous function $\beta:\Real\to \Real$ is said to be an extended class $\mathcal{K}_{\infty}$ function if $\beta$ is strictly monotonically increasing and $\beta(0) = 0$, $\lim_{x\to \infty}\beta(x) = \infty$. For $x\in \Real^n$, Lie derivative of a scalar function $S(x) \in \Real$ along a vector field $h(x)\in \Real^n$ or matrix field $h(x) \in \Real^{n\times m}$ is represented by $L_h S(x) = \frac{\partial S^{\Tran}}{\partial x} h(x)$.

\section{Mathematical Preliminaries}\label{Section_Pre}

\subsection{System Model and Problem Formulation}\label{Section_Pre_sub_SMPF}
In this paper, the control affine system is considered as
\begin{align}
	\dot x(t) = f(x(t)) +  g(x(t))u(t) +  \theta\Delta(x(t),u(t)) + d(t) , \label{system_dynamics}
\end{align}
where $x(t)\in \Real^n$, $u(t)\in\mathcal{U}\subset \Real^m$ and $\theta\in\Theta \subset \Real^{n\times p}$ are system state, control input in admissible set $\mathcal{U}$ and some constant, bounded unknown parameters in a known compact convex set $\Theta$, respectively. The nominal dynamics $f:\Real^n \to \Real^n$ and control input dynamics $g:\Real \to \Real^{n \times m}$ are assumed to be locally Lipschitz, $\Delta: \Real^{n+m} \to \Real^{p}$ is measurable regressor matrix while $d: \Real \to \Real^{n}$ consists of some unknown disturbances assumed to be bounded as $\norm{d(t)}\leq \bar d$, $\forall t > 0$.
The main problems studied in this work are: (a) under what criteria can have the unknown parameters estimated in a finite-time; (b) how to satisfy these criteria and control performance without violation of safety; (c) how to describe the robustness or impact of the unpleasant disturbances. Some inspiring results are reviewed as follows to help readability.

\subsection{Finite-Time Parameter Identification}\label{Section_Pre_sub_FTPI}
The unknown parameters can be estimated using collected data sampled from a control process. The finite-time parameter estimation can be realized by means of dynamic regressor extension and mixing (DREM) \cite{matrix_estimation_finite_time,finite_time_robust}, which is briefly discussed as follows. For a given linear regression equation (LRE) as
\begin{equation}
	y(t) = z^T(t) \phi + w(t), \label{pre_LRE_origin}
\end{equation}
$y\in \Real$ and $z\in \Real^{p}$ are measurable quantities, $w(t)\in \Real$ is the disturbance, the unknown parameters $\phi \in \Real^q$ can be estimated by $\hat \phi \in \Real^q$ via an appropriate updating law. To weaken the need of persistent excitation on signal $ z(t)$,  the $\mathcal{L}_\infty$-stable linear operators $\mathcal{H}:\Real\to \Real^{q}$ such as linear time-invariant filters with transfer function $\mathcal{H}_G(s) = col(\frac{p_1}{s + \lambda_1}, \dots,\frac{p_q}{s + \lambda_q})$ are introduced to expand vector $ z(t)$ to square matrix $Z(t) = \left[\mathcal{H}(z_1(t)),\dots,\mathcal{H}(z_q(t))\right]\in \Real^{q\times q}$ where $z_i(t)$ represents the $i$th element of $z(t)$. To further decouple the estimation process, the adjoint matrix $adj\left\{ Z(t)\right\}$ of $Z(t)$ is utilized with the fact that $adj\left\{ Z(t)\right\} Z(t) = \delta(t) I_q$, where $\delta(t) = det(Z(t))$. Correspondingly, LRE \eqref{pre_LRE_origin} is transformed into 
\begin{equation}
	Y(t) = \delta(t) \phi + W(t), \label{pre_LRE_decoupled}
\end{equation}
where $Y = adj\left\{ Z(t)\right\}\mathcal{H}(y)$ and $W = adj\left\{ Z(t)\right\}\mathcal{H}(w)$. It can be seen that $\phi$ in \eqref{pre_LRE_decoupled} can be estimated in an element-wise manner since $\delta(t)$ is a scalar. Consider each row in \eqref{pre_LRE_decoupled} as $Y_i(t) = \delta(t) \phi_i + W_i(t)$, the estimator $\hat \phi$ is updated following
\begin{equation}
	\dot {\hat\phi}_i (t) = \gamma \delta(t) \left(Y_i(t) - \delta(t) \hat\phi(t) \right)^r, \quad r \in (0,1]\label{pre_LRE_update_law}
\end{equation}
where $\gamma > 0$ is the learning rate. The following results show the stability and robustness of the updating law \eqref{pre_LRE_update_law}.
\begin{proposition}[\cite{finite_time_robust}]
	\label{prop_finitetime}
	If $\delta(t) \notin \mathcal{L}_2$, i.e. $\int_0^\infty \delta^2(t) \, dt = +\infty$, then under the law \eqref{pre_LRE_update_law}, one has
	\begin{itemize}[\IEEEsetlabelwidth{Z}]
		\item[(A)] If $r = 1$ and $\norm{W} = 0$, the estimator $\hat\phi$ asymptotic convergences to $\phi$ monotonically;
		\item[(B)] If $r\in(0,1)$ and $\norm{W} = 0$, the estimator $\hat\phi$ convergences to $\phi$ monotonically in finite-time;
		\item[(C)] If $r\in(0,1)$ and $\norm{W} \leq \bar W$, the error system $e(t) = \phi - \hat\phi(t)$ is short-finite-time input-to-state stable. 
	\end{itemize}
\end{proposition}
In Proposition \ref{prop_finitetime}, the proof of (A) can be seen in \cite{DREM_begin}, the proofs of (B) and (C) can be seen in \cite{finite_time_robust}. Another finite-time algorithm of DREM is proposed in \cite{matrix_estimation_finite_time}, in which a scalar signal is stored to dynamically adjust learning rate $\gamma$. Compared with \cite{finite_time_robust}, the injection of high-gain is eliminated. However, the impact of disturbance has not been researched theoretically in \cite{matrix_estimation_finite_time}. While the unknown parameters are tuned according to collective samples of randomly given control input and system state, security issues are often neglected. An effective method for safety control is presented as follows.

\subsection{Adaptive Control Barrier Function}\label{Section_Pre_sub1_RCBF}
The system is safe if $\exists t_0$ such that $x(t_0)\in \mathcal{C}_{is}$, there is $x(t) \in \mathcal{C}_{is}$, $\forall t \ge t_0$, where $\mathcal{C}_{is}$ is a forward invariant set, which can be determined by a superlevel set $\mathcal{C}=\{x \in \Real^n \vert ~B(x) \ge 0\} $ of a CBF function $B(x)$. In addition, the CBFs considered in this paper are assumed to has uniform relative degree one.
\begin{definition}[CBF \cite{cbfdefine}]
	\label{definition_CBF}
	For a closed convex set $\mathcal{C}$, a continuously differentiable function $B(x):\Real^n \to \Real$ is a zeroing CBF on $\mathcal{C}$ for a system $\dot x = h(x,u,d)$ if for all $x\in \mathcal{C}$,
	\begin{equation}
		\sup_{u\in \mathcal{U}} \left[\frac{\partial B}{\partial x}(x) h(x,u,d)\right] \ge -\alpha(B(x)), \label{definitionCBF}
	\end{equation}
	where $\alpha: \Real \to \Real$ is an extended class $\mathcal{K}_{\infty}$ function. 
\end{definition}
\begin{lemma}[\cite{cbfdefine}]
	\label{lemma_ForwardInvariant}
	Let $B(x)$ be a CBF on a closed convex set $\mathcal{C}=\{x~\vert~B(x)\ge 0\}$, then any Lipschitz continuous controller satisfying \eqref{definitionCBF} renders safety on the set $\mathcal{C}$, i.e. $\mathcal{C}$ is forward invariant.
\end{lemma}
In \eqref{definitionCBF}, if the disturbances $d$ are unknown, i.e. the dynamics $h(x,u,d)$ is uncertain, one can not verify whether a control input results in safety. Same problem exists with the unknown parameters if considering system \eqref{system_dynamics}. To deal with this, a tightened set can be designed as $\mathcal{C}_\theta \subseteq \mathcal{C}$ to reserve robustness. Before the adaptive CBFs are given, it is assumed for simplicity that $d(t)\equiv 0$. It is assumed after \eqref{system_dynamics} that the uncertain parameters are in a compact known set $\Theta$, then the worst-case estimation error bound can be generated as $\tilde{\theta}_{\max} ^\Tran \tilde{\theta}_{\max}$, where $\tilde{\theta}_{\max} = \sup\{ \hat \theta(t) - \theta\}$ and $\hat \theta(t)$ is the estimated virsion of the unknown parameters. With the adaptive CBF $B_a(x,\theta)$, take $\mathcal{C}_\theta=\{x \in \Real^n~\vert ~B_a(x,\theta) \ge \tilde{\theta}_{\max} ^\Tran \tilde{\theta}_{\max} \} $. The criteria for safety can be seen in \cite{RaCBF_IEEECSL} [Theorem 2]. Note that the unknown error bound or estimation set $\Theta(t)$ is shrinking through adaptation, the tightened set $\mathcal{C}_\theta$ is approaching to $\mathcal{C}$ when $\tilde \theta_{\max}(t) \to 0$, a brief discussion is presented in \cite{RaCBF_IEEECSL} [Theorem 3]. Following the same idea, a fixed time adaptation law combining with adaptive CBF is studied in \cite{fixedtimeACBF}. The statements therein are given below.
\begin{proposition}[\cite{fixedtimeACBF}]
	\label{prop_2}
	For time-varying worst-case error bound $\tilde{\theta}_{\max}(t) = \sup \{\hat \theta(t) - \theta \}$, the set $C_\theta = \{ x\in \Real^n ~\vert~ B_a(x,\theta) \ge \tilde{\theta}_{\max}^\Tran(t)\tilde{\theta}_{\max}(t)\}$ is forward invariant, if
	\begin{align}
		&\sup_{u \in \mathcal{U}} \left\{ L_f B_a(x,\theta) + L_g B_a(x,\theta) u + \Psi\right\}\nonumber\\
		\ge & -\alpha(B_a(x,\theta) - \tilde{\theta}_{\max}^\Tran(t)\tilde{\theta}_{\max}(t)) + \tilde{\theta}_{\max}^\Tran(t)\dot{\tilde{\theta}}_{\max}(t), \label{prop_ACBF}
	\end{align}
	in which $\Psi$ is the lower bound of $ L_\theta B(x,\theta) \Delta(x)$, namely $\Psi \leq L_\theta B(x,\theta) \Delta(x)$.
\end{proposition}
The improvement of \cite{fixedtimeACBF} compared to \cite{RaCBF_IEEECSL} is the relaxed adaptation law. By taken $B_a(x,\theta) = B(x) - \tilde{\theta}_{\max}^\Tran(t)\tilde{\theta}_{\max}(t)$, it is known that $B_a(x,\theta)$ is a CBF by Definition \ref{definition_CBF}, and $\mathcal{C}_\theta$ is sufficiently to be forward invariant when \eqref{prop_ACBF} is satisfied according to Lemma \ref{lemma_ForwardInvariant}.

\section{Finite-Time Parameter Estimation with Safety}\label{Section_ALsafe}
In this section, a finite-time parameter estimation algorithm for nonlinear systems \eqref{system_dynamics} improved from DREM is studied. The algorithm is combined with CBF method to ensure system safety with input constraints. Without loss of generality, assume $d(t) \equiv 0$ in this section. The impact of the disturbances $d(t)$ will be analyzed in the next section. 

\par The LRE transformed form \eqref{system_dynamics} is given as
\begin{align}
	X(x, u)=  \theta\Delta(x,u) , \label{system_regression}
\end{align}
where $X(x,u) = \dot x(t) - f(x(t)) -  g(x(t))u(t)$. It is assumed that $X(x,u)$ and $\Delta(x,u)$ is measurable. The derivative $\dot x$ can be approximated by $\left(x(t) - x(t + \Delta t)\right) / \Delta t$, which may introduce approximation error $\tilde{d}(t)$ into LRE \eqref{system_regression}. This error can be viewed as new disturbances which can be added into $d(t)$ because the Lipschitz continuity of the system dynamics ensures that $\norm{\tilde{d}(t)}$ is bounded. Alternatively, linear filters can be used to obtain the new LRE as shown in \cite{PANTELEY20021125} [Lemma 3.1].
\par Different from \eqref{pre_LRE_origin} where $\phi$ is in a vector field, $\theta$ of \eqref{system_regression} is in a matrix field. This case has been studied in \cite{matrix_estimation_finite_time}, where an adaptive controller has also been introduced to ensure stability. The same process is followed below with a different finite-time estimation algorithm.
\par For linear $\mathcal{L}_\infty-$stable operators $\mathcal{H}:\Real \to \Real^q$, applying them on the transpose form of \eqref{system_regression}, one has
\begin{align}
	\mathcal{X}_H= \mathcal{Z}_H \theta^\Tran, \label{DREM_1}
\end{align}
in which $\mathcal{X}_H = \left(\mathcal{H}(X_1), \dots ,\mathcal{H}(X_n)\right) \in \Real^{q \times n}$ and $\mathcal{Z}_H = \left(\mathcal{H}(\Delta_1), \dots ,\mathcal{H}(\Delta_q)\right) \in \Real^{q \times q}$ is a square matrix. Multiply $adj\{\mathcal{Z}_H\}$ on the left-side of \eqref{DREM_1} yields
\begin{align}
	\mathcal{X}= \delta(t)\theta^\Tran, \label{DREM_2}
\end{align}
where $\mathcal{X} = adj\{\mathcal{Z}_H\}\mathcal{X}_H$ and $\delta(t) = det\{\mathcal{Z}_H\}$. The significance of \eqref{DREM_2} is to obtain the element-wise LRE that will benefit the relaxation of PE condition and the design of finite-time estimation algorithm. Let $\hat \theta (t)$ be the parameters as estimation of $\theta$. Considering each element of the extended LRE \eqref{DREM_2}, i.e. $\mathcal{X}_{ij}= \delta(t)\theta_{ji}$, a high-gain estimation dynamics similar to \eqref{pre_LRE_update_law} is presented as
\begin{equation}
	\dot {\hat\theta}_{ij} (t) = \gamma \delta(t) \left(\mathcal{X}_{ji}(t) - \delta(t) \hat\theta_{ij}(t) \right)^r, \quad r \in (0,1).\label{finitetime_law}
\end{equation}
The element-wise estimation error is denoted by $\tilde \theta_{ij}(t) = \hat\theta_{ij}(t) - \theta_{ij}$, $\theta_{ij}\in \Theta_{ij}$, $i=1,\dots,n$ and $j = 1,\dots p$. Then with \eqref{DREM_2}, the error dynamics is
\begin{equation}
	\dot {\tilde\theta}_{ij} (t) = -\gamma \delta(t) \left(\delta(t)\theta_{ij} - \delta(t) \hat\theta_{ij}(t) \right)^r.\label{finitetime_error_dynamics}
\end{equation}
Before considering safety and constraints, the theoretical analysis of finite-time algorithm \eqref{finitetime_law} is given below.
\begin{assumption}
	\label{assume_IEcondition}
	A signal $\phi(t)$ is interval excitation (IE) in $(0,t_c)$, $t_c> 0$ if it satisfies $
		\int_0^{t_c} \phi(t)\phi^\Tran(t) \, dt \ge \beta I,
	$
	where $\beta > 0$ is a known constant.
\end{assumption}
\begin{proposition}
	\label{proposition_FTnoD}
	Consider the updating law \eqref{finitetime_law} and $d\equiv 0$. If signal $\vert\delta(t)\vert^{\frac{1 + r}{2}}$ is IE within an interval $(0, t_c)$, i.e. satisfies Assumption \ref{assume_IEcondition}, then element-wise estimation error $\tilde \theta_{ij}(t)$ monotonically converges to zero within $(0, t_c)$ as long as
	\begin{align}
		\gamma \ge \frac{2\bar\theta_{ij}^2}{(1-r)\beta},\quad \bar \theta_{ij} = \sup_{\hat \theta_{ij} \in \Theta_{ij}} \left\{ \vert\hat \theta_{ij} - \theta_{ij}\vert \right\}. \label{prop_3_learningrate}
	\end{align}
\end{proposition}
\textbf{Proof.} Construct a Lyapunov candidate $V(\tilde \theta_{ij}(t)) = \frac{1}{2} {\tilde\theta}_{ij}^2(t) $ whose derivative of $t$ is
\begin{align}
	\dot V(\tilde\theta_{ij}(t)) = & -\gamma \delta(t) \tilde{\theta}_{ij}(t)\left(\delta(t)\theta_{ij} - \delta(t) \hat\theta_{ij}(t) \right)^r\nonumber \\
	= & -\gamma \vert\delta(t)\vert^{1 + r} \vert\tilde{\theta}_{ij}(t)\vert^{1 + r} \nonumber\\
	= & -\gamma \vert\delta(t)\vert^{1 + r} \left[\tilde{\theta}_{ij}(t)^{2}\right]^{\frac{1 + r}{2}}\nonumber\\ 
	= & - \kappa^2(t) V^{\frac{1 + r}{2}}(\tilde\theta_{ij}),
\end{align}
where $\kappa(t) = \vert\delta(t)\vert^{\frac{1 + r}{2}}$. Solving the equation above, one has
\begin{align}
	V(\tilde \theta_{ij}(t)) = \left[V(\tilde{\theta}_{ij}(0))-\frac{(1 - r)\gamma}{2}\int_{0}^t \kappa^2(t)\, dt  \right]^{\frac{2}{1-r}}. \label{prop_3_errorbound}
\end{align}
Under Assumption \ref{assume_IEcondition}, $\forall t \ge t_c$, $\int_0^t \kappa^2(t) \ge \beta$ with the fact that $\kappa^2(t) \ge 0$. Then, for $t=t_c$, there is
\begin{align*}
	V(\tilde \theta_{ij}(t)) \leq  \left[-\frac{(1 - r)\gamma\beta}{2} + V(\tilde{\theta}_{ij}(0)) \right]^{\frac{2}{1-r}},
\end{align*}
which implies $V(\tilde \theta_{ij}(t))$ converges to zero within $(0,t_c)$ if $\gamma \ge \frac{2V(\tilde{\theta}_{ij}(0))}{(1-r)\beta}$, which is reflected in \eqref{prop_3_learningrate}. The proof is done.

\par To achieve finite-time estimation, IE condition should be satisfied. This is always implemented by randomly selected noise signals \cite{finite_time_robust,DREM_begin} or elaborately designed control input \cite{matrix_estimation_finite_time}. These methods work in simulations or simple scenarios needless to consider constraints or conflicts. However, control systems \eqref{system_dynamics} always has constraints e.g. input constraints, as well as conflicts such as performance trade-off or safety \cite{cbfdefine}. Especially for safety, the collective data sampled from a controlled trajectory should be in a safe region. However, the estimated parameters of system dynamics are (partially) unknown, bringing a dilemma of how to design the controller to ensure safety. As the unknown parameters may be the backbone of system dynamics, it is not perfect to consider them as big disturbances and use some worst-case robust controller \cite{RaCBF_IEEECSL}. Luckily, with the finite-time adaptation law, the evolution of estimation error can be formulated, as shown in Proposition \ref{proposition_FTnoD}. With these observations, a control policy leveraging adaptive CBF for safe adaptation is given below. 

\begin{theorem}
	\label{theorem1}
	Consider a superlevel set $\mathcal{C}_{\hat \theta} = \{x\in \Real^{n} \vert B(x) > 0\}$ of a CBF $B(x)$ with
	\begin{align}
		\sup_{u \in \mathcal{U}} &\left\{ L_f B(x) + L_g B(x) u + L_{\hat \theta} B(x)\Delta (x,u)\right.\nonumber\\
		& \left. - \psi(t)\right\} \ge -  \alpha(B(x)),\label{th1_acbf}
	\end{align}
	where $\alpha$ is an extended class $\mathcal{K}_\infty$ function and 
	\begin{equation}
		\psi(t) = \sum_{i=1}^n \sum_{j=1}^p \tilde\theta^r_{ij}(t)\vert \frac{\partial B(x)}{\partial x_i} \Delta_{j}(x,u)\vert, \label{th1_cbf_adaptive_item}
	\end{equation}
	in which $\tilde \theta^r_{ij}(t) = \left[ \Delta\theta_{ij}^2 -\frac{(1 - r)\gamma}{2}\int_{0}^t \kappa^2(t)\, dt  \right]^{\frac{1}{1-r}}$. Then for any locally Lipschitz continuous controller satisfying \eqref{th1_acbf} renders $\mathcal{C}_{\hat \theta}$ forward invariant with finite-time adaptation law \eqref{finitetime_law} and learning rate \eqref{prop_3_learningrate}. Moreover, $\psi(t)\to 0$ within interval $(0,t_c)$ as long as signal $\kappa(t) = \vert\delta(t)\vert^{\frac{1 + r}{2}}$ is IE within an interval $(0, t_c)$, $\exists t_c > 0$. Then $\mathcal{C}_{\hat \theta}$ becomes the safe set of original systems \eqref{system_dynamics} as $\hat \theta(t)=\theta$, $\forall t > t_c$.
\end{theorem}
\textbf{Proof.} The convergence of finite-time parameter estimation has been given in Proposition \ref{proposition_FTnoD}. In what follows, the IE condition is considered to be satisfied, i.e. $\tilde{\theta}_{ij} \to 0$ within $(0,t_c)$. Note that \eqref{prop_3_errorbound} describes the element-wise upper bound of estimation error which is monotonically non-increasing. Since $\theta_{ij}$ is unknown, the worst case of initial error is denoted as $\Delta\theta_{ij} = \sup_{\theta_a,\theta_b \in \Theta_{ij}}\left\{ \vert \theta_a - \theta_b \vert \right\}$. Introducing $\Delta\theta_{ij}$ into \eqref{prop_3_errorbound}, an upper bound is obtained as
\begin{align}
	\vert \tilde \theta_{ij}(t)\vert \leq \left[ \Delta\theta_{ij}^2 -\frac{(1 - r)\gamma}{2}\int_{0}^t \kappa^2(t)\, dt  \right]^{\frac{1}{1-r}}. \label{th1_upperboundij}
\end{align}
For a zeroing CBF $B(x)$, \eqref{definitionCBF} is determined as
\begin{align}
	&\sup_{u \in \mathcal{U}} \left\{ L_f B(x) + L_g B(x) u + \frac{\partial B}{\partial x}(x) \theta \Delta(x,u) \right\}\nonumber\\
	= & \sup_{u \in \mathcal{U}} \left\{ L_f B(x) + L_g B(x) u + \frac{\partial B}{\partial x}(x) \left(\hat\theta - \tilde \theta\right) \Delta(x,u) \right\}\nonumber\\
	= & \sup_{u \in \mathcal{U}} \left\{ L_f B(x) + L_g B(x) u + L_{\hat\theta} B(x)\Delta(x,u) - \xi(t) \right\}\nonumber\\
	\ge& -\alpha(B(x)), \label{th1_proof_cbf}
\end{align}
where $\xi(t) = \frac{\partial B}{\partial x}(x) \tilde \theta\Delta(x,u)$. Any locally Lipschitz continuous controller that satisfies \eqref{th1_proof_cbf} renders $\mathcal{C}_{\hat \theta}$ forward invariant as shown in Lemma \ref{lemma_ForwardInvariant}. With the definition of $\tilde{\theta}(t)$ and \eqref{th1_upperboundij}, one has
\begin{align}
	\xi(t) = &\frac{\partial B}{\partial x}(x) \tilde \theta\Delta(x,u)\nonumber\\
	=& \sum_{i=1}^n \sum_{j=1}^p \tilde\theta_{ij}(t) \frac{\partial B(x)}{\partial x_i} \Delta_{j}(x,u)\nonumber\\
	\leq & \sum_{i=1}^n \sum_{j=1}^p \vert\tilde\theta_{ij}(t)\vert \vert\frac{\partial B(x)}{\partial x_i} \Delta_{j}(x,u)\vert.
\end{align}
Therefore, when \eqref{th1_cbf_adaptive_item} is satisfied, \eqref{th1_proof_cbf} will hold. Then $\mathcal{C}_{\hat\theta}$ is forward invariant for any Lipschitz continuous control input $u$ satisfying \eqref{th1_cbf_adaptive_item}. When estimation is done, the upper bound of $\tilde{\theta}_{ij}$ reduces to zero, i.e. $\vert \tilde{\theta}(t) \vert = 0$, $\forall t > t_c$, one has $\psi(t) = 0$. Then \eqref{th1_cbf_adaptive_item} is equal to \eqref{th1_proof_cbf} and $\hat\theta (t)=\theta$. As a results, $\mathcal{C}_{\hat\theta}$ becomes the safe set of known systems \eqref{system_dynamics} without uncertain estimation error.

\par The difference between Theorem \ref{theorem1} and Proposition \ref{prop_2} is the definition of CBF. By introducing a tightened safe set $\mathcal{C}^s_{\hat\theta}\subset\mathcal{C}_{\hat\theta}$. As stated in \cite{RaCBF_IEEECSL}, a tightened set provides a less conservative safe constraint than \eqref{th1_acbf}. This algorithm is presented below as a comparison of Theorem \ref{theorem1}.
\begin{corollary}
	\label{corolarry1}
	Use the same notations in Theorem \ref{theorem1}. Consider a subset of $\mathcal{C}_{\hat\theta}$ defined by $\mathcal{C}^s_{\hat\theta} = \left\{ x\in \Real^n \vert B_(x) \ge \Xi(t) \right\}$ of a CBF $B(x)$ satisfying
	\begin{align}
		&\sup_{u \in \mathcal{U}} \left\{ L_f B(x) + L_g B(x) u + L_{\hat \theta} B(x)\Delta (x,u)\right.
		\left. - \psi(t)\right\}\nonumber\\
		\ge & -  \alpha\left(B(x) - \Xi(t)\right) + \sum_{i=1}^n \sum_{j=1}^p \tilde\theta^r_{ij}(t) \dot {\tilde\theta}^r_{ij}(t) ,\label{coro1_acbf}
	\end{align}
	where $\Xi(t) = \sum_{i=1}^n \sum_{j=1}^p \tilde\theta_{ij}^{r2}(t) $. Then for any locally Lipschitz continuous controller satisfying \eqref{th1_acbf} renders $\mathcal{C}_{\hat \theta}$ forward invariant with finite-time adaptation law \eqref{finitetime_law} and learning rate \eqref{prop_3_learningrate}.
\end{corollary}
\textbf{Proof.} Let $B_a(x) = B(x) - \Xi(t)$. Calculate its derivative of $t$, one has
\begin{align}
	\dot B_a(x) = & L_f B(x) + L_g B(x) u + L_{\hat \theta} B(x)\Delta (x,u)\nonumber\\
	& + \xi(t) - \sum_{i=1}^n \sum_{j=1}^p \tilde\theta^r_{ij}(t) \dot {\tilde\theta}^r_{ij}(t).
\end{align}
When \eqref{coro1_acbf} is satisfied, there is
\begin{align*}
	\sup_{u\in \mathcal{U}} \dot B_a(x) &+ \sum_{i=1}^n \sum_{j=1}^p \tilde\theta_{ij}^r(t) \dot {\tilde\theta}^r_{ij}(t) \nonumber\\
	\ge & -\alpha(B_a(x)) + \sum_{i=1}^n \sum_{j=1}^p \tilde\theta^r_{ij}(t) \dot {\tilde\theta}^r_{ij}(t).
\end{align*}
Subtracting the same items, one has $\sup_{u\in \mathcal{U}} \dot B_a(x) \ge  -\alpha(B_a(x))$ i.e. $B_a(x)$ is a CBF of set $C_{\hat\theta}^s = \left\{ B_a(x) > 0\right\}$. Therefore, $C_{\hat\theta}^s$ is forward invariant. It is noted that $\Xi(t) = 0$ $\forall t>t_c$, thus $B_a(x(t)) = B(x(t))$, which implies $C_{\hat\theta}^s$ will be the safe set of origin systems \eqref{system_dynamics} benefited from the finite-time adaptation.

\par A more discussion on the two algorithms is given. Notice that the left part of \eqref{th1_acbf} and \eqref{coro1_acbf} is the same, only the right side of the safe constraints needs to be analyzed. The two different parts are listed below which are recognized as "Sbound1" for \eqref{th1_acbf} and "Sbound2" for \eqref{coro1_acbf}, respectively.
\begin{align*}
	\begin{cases}
		\text{Sbound1:} & -  \alpha(B(x)),\\
		\text{Sbound2:} & -\alpha\left(B(x) - \Xi(t)\right) + \sum_{i=1}^n \sum_{j=1}^p \tilde\theta^r_{ij}(t) \dot {\tilde\theta}^r_{ij}(t).
	\end{cases}
\end{align*}
Recall the definition of CBF, a lower bound is always preferred for relaxation \cite{cbfdefine}. Since the estimation error of each unknown parameter is monotonically non-increasing, one has $\tilde\theta^r_{ij}(t) \dot {\tilde{\theta}}^r_{ij}(t) \leq 0$. Therefore, choosing Sbound1 or Sbound2 is highly dependent on the real-time measurement or calculation of $B(x)$, $\tilde{\theta}(t)$ and $\dot {\tilde{\theta}}_{ij}(t)$. For instance, consider $B(x)=1$, $\alpha(x) = x^2$ and one unknown parameter with estimation error $\tilde{\theta} = 0.5$. It is obtained that Sbound1$=-1.0$, Sbound2$=-0.5625+0.5\dot{\tilde{\theta}}$. If $\dot{\tilde{\theta}}(t) = 0$, i.e. $\kappa(t)=0$, Sbound1 is chosen; when $\dot{\tilde{\theta}}(t) < -1$, Sbound2 is better. As a result, a new algorithm with switched safe constraints can be delivered.
\begin{corollary}
	\label{corolarry2}
	Use the same notations in Theorem \ref{theorem1} and Corollary \ref{corolarry1}. Consider a subset of $\mathcal{C}_{\hat\theta}$ defined by $\mathcal{C}^s_{\hat\theta} = \left\{ x\in \Real^n \vert B_(x) \ge \Xi(t) \right\}$ of a CBF $B(x)$ satisfying
	\begin{align}
		&\sup_{u \in \mathcal{U}} \left\{ L_f B(x) + L_g B(x) u + L_{\hat \theta} B(x)\Delta (x,u)\right.
		\left. - \psi(t)\right\}\nonumber\\
		\ge & \min\left( \text{Sbound1},\text{Sbound2} \right).\label{coro2_acbf}
	\end{align}
	Then for any locally Lipschitz continuous controller satisfying \eqref{th1_acbf} renders $\mathcal{C}_{\hat \theta}$ forward invariant with finite-time adaptation law \eqref{finitetime_law} and learning rate \eqref{prop_3_learningrate}.
\end{corollary}
The proof is the same as Theorem \ref{theorem1} and Corollary \ref{corolarry1}, which is omitted for brevity.
\begin{remark}
	As discussed above, the conservativeness is reduced in Corollary \ref{corolarry2}. The tightened safe set proposed in \cite{RaCBF_IEEECSL} is also used in this algorithm. Differently, the worst-case error bounds of unknown parameters are decreasing monotonically and reduce to zero in a finite time such that the conservativeness of tightened safe set will approach to the origin safe set during the adaptation process. Note that the same idea was presented in \cite{fixedtimeACBF} to obtain the time-varying error bound and its derivative, as stated in Proposition \ref{prop_2}. In addition to the relaxation of conservativeness as discussed above compare to Proposition \ref{prop_2}, the element-wise error bounds introduced from DREM is much simple than the filter-based estimation error of entire parameter space $\Theta$. Moreover, it will be seen in the next section that the robustness of DREM can be theoretically analyzed however the identification algorithm in \cite{fixedtimeACBF} may not be robust to disturbances. Same issue exists in the estimation algorithm in \cite{safe_uncertain_ELsystem}.
\end{remark}
\begin{remark}
	It should be noted that the selected finite-time DREM implies the high gains, which could discount the stability in control scenario as discussed in \cite{DREM_improve_time_varying}. An Alternative finite-time DREM method has been presented in \cite{matrix_estimation_finite_time} with elaborate control design. This method can be directly planted in the presented safe adaptation algorithms. Nevertheless, the robustness of this finite-time DREM method needs to be further studied.
\end{remark}
\section{Robustness and Algorithm Design}\label{Section_AL_robust}
It is assumed earlier that no disturbance is included in systems \eqref{system_dynamics}. Unfortunately, the disturbances are unavoidable and unmeasurable as well. The robustness of finite-time DREM \eqref{finitetime_law} has been analyzed in \cite{finite_time_robust} for unknown parameters in vector field. Same process is utilized in this paper for unknown parameters in matrix field. Next, a robust CBF is constructed considering the bounded disturbances. For goal-oriented controller, a robust CLF constraint is added to the algorithm. Finally, a point-wise algorithm based on quadratic programming is given.
\begin{definition}[short-finite-time ISS \cite{finite_time_robust}]
	The estimation process \eqref{finitetime_error_dynamics} is said to be short-finite-time ISS if for all $\tilde\theta_{ij}(0)\in \Theta_{ij}$ and bounded disturbances $\norm{w(t)}\leq \bar w$,
	\begin{equation}
		\vert{\tilde{\theta}_{ij}(t)}\vert \leq \alpha(\tilde\theta_{ij}(0),t) + \beta(\norm{w}), \quad \forall t\in\left[0,t_c\right]
	\end{equation}
	where $\alpha(s,t)$ is a class $\mathcal{KL}$ function with $\alpha(s,t) = 0$, $\forall t>t_c$, $\beta$ is a class $\mathcal{K}$ function.
\end{definition}
\begin{lemma}[\cite{finite_time_robust}]
	\label{lemma_finite-time-robust}
	With the finite-time DREM \eqref{finitetime_law}, the element-wise error system $\mathcal{X}_{ji}= \delta(t)\theta_{ij} + w(t)$ is short-finite-time ISS with bounded disturbance $\norm{w}\leq \bar w$.
\end{lemma}
The main process is to show that adding disturbances $d(t)$ in \eqref{system_dynamics} results in a modified version of LRE \eqref{DREM_2} which takes the form as Lemma \ref{lemma_finite-time-robust} requires. With $d(t)$, \eqref{system_regression} becomes
\begin{align*}
	X(x, u)=  \theta\Delta(x,u) + \hat d(t),
\end{align*}
where $\hat d(t)=d(t) + d^\prime (t)$ and $d^\prime(t)$ represents the approximation error of the derivative $\dot x$ with $\norm{d^\prime(t)}\leq \bar d^\prime$. Therefore, $\norm{\hat d(t)}\leq \bar d + \bar d^\prime$. Same operations $\mathcal{H}$ applied can obtain
\begin{align*}
	\mathcal{X}_H= \mathcal{Z}_H \theta^\Tran + \mathcal{D}_H,
\end{align*}
where $\mathcal{D}_H = \left(\mathcal{H}(\hat d_1), \dots ,\mathcal{H}(\hat d_n)\right) \in \Real^{q \times n}$. Using the linear operators implies $\norm{\mathcal{D}_H} \leq \bar{\mathcal{D}}_H(\bar d + \bar d^\prime)$, where the upper bound $\bar{\mathcal{D}}_H$ is dependent on the bounds of disturbances and $\mathcal{H}$.
\begin{align*}
	\mathcal{X}= \delta(t)\theta^\Tran + \mathcal{D},
\end{align*}
where $\mathcal{D} = adj\{\mathcal{Z}_H\}\mathcal{D}_H$. An element-wise LRE takes the form as
\begin{align}
	\mathcal{X}_{ji}= \delta(t)\theta_{ij} + \mathcal{D}_{ji}. \label{elementwise_lre_robust}
\end{align}
\begin{assumption}
	\label{assumption_bound_robust}
	The disturbances $\mathcal{D}_{ji}$ in \eqref{elementwise_lre_robust} are bounded by $\norm{\mathcal{D}_{ji}}\leq \bar{\mathcal{D}}_{ji}$, $i = 1,\dots,n$, $j=1,\dots,q$.
\end{assumption}
\begin{proposition}
	\label{proposition_FTrobust}
	Use the same notations in Proposition \ref{proposition_FTnoD} and let Assumption \ref{assumption_bound_robust} hold. Consider the updating law \eqref{finitetime_law}. If signal $\vert\delta(t)\vert^{\frac{1 + r}{2}}$ is IE within an interval $(0, t_c)$, i.e. satisfies Assumption \ref{assume_IEcondition}, then element-wise estimation error system \eqref{finitetime_error_dynamics} is short-finite-time ISS in $(0, t_c)$ as long as $
		\gamma \ge \frac{2\bar\theta_{ij}^2}{(1-r)\beta}$.
\end{proposition}
As the disturbances $d(t)$ is unmeasurable, it is necessary to consider the worst-case impact in CBF \cite{robust_cbf}. 
\begin{theorem}
	Use the same notations in Theorem \ref{theorem1} and Corollary \ref{corolarry1}. Consider a subset of $\mathcal{C}_{\hat\theta}$ defined by $\mathcal{C}^s_{\hat\theta} = \left\{ x\in \Real^n \vert B_(x) \ge \Xi(t) \right\}$ of a robust adaptive CBF $B(x)$ satisfying
	\begin{align}
		&\sup_{u \in \mathcal{U}} \left\{ L_f B(x) + L_g B(x) u + L_{\hat \theta} B(x)\Delta (x,u)\right.
		\left. - \psi(t)\right\}\nonumber\\
		\ge & \min\left( \text{Sbound1},\text{Sbound2} \right) - \norm{\frac{\partial B}{\partial x}(x) }\bar d.\label{thm2_racbf}
	\end{align}
	Then for any locally Lipschitz continuous controller satisfying \eqref{th1_acbf} renders $\mathcal{C}_{\hat \theta}$ forward invariant.
\end{theorem}
\textbf{Proof.} The proof is presented for the case of choosing Sbound2. The other case of choosing Sbound1 is omitted for brevity. Let $B_a(x) = B(x) - \Xi(t)$. Calculate its derivative of $t$, one has
\begin{align}
	\dot B_a(x) = & L_f B(x) + L_g B(x) u - L_{\hat \theta} B(x)\Delta (x,u) \nonumber\\
	& + \frac{\partial B}{\partial x}(x) d(t) + \xi(t) - \sum_{i=1}^n \sum_{j=1}^p \tilde\theta^r_{ij}(t) \dot {\tilde\theta}^r_{ij}(t).
\end{align}
Notice that $\frac{\partial B}{\partial x}(x) d(t) \ge -\norm{\frac{\partial B}{\partial x}(x) d(t)} \ge -\norm{\frac{\partial B}{\partial x}(x) }\bar d$. When \eqref{thm2_racbf} is satisfied, there is
\begin{align*}
	&\sup_{u\in \mathcal{U}} \dot B_a(x) + \sum_{i=1}^n \sum_{j=1}^p \tilde\theta_{ij}^r(t) \dot {\tilde\theta}^r_{ij}(t) +\norm{\frac{\partial B}{\partial x}(x) }\bar d\nonumber\\
	\ge & \sup_{u \in \mathcal{U}} \left\{ L_f B(x) + L_g B(x) u + L_{\hat \theta} B(x)\Delta (x,u)\right.
	\left. - \psi(t)\right\}\nonumber\\
	\ge & -\alpha(B_a(x)) + \sum_{i=1}^n \sum_{j=1}^p \tilde\theta^r_{ij}(t) \dot {\tilde\theta}^r_{ij}(t) +\norm{\frac{\partial B}{\partial x}(x) }\bar d.
\end{align*}
Subtracting the same items, one has $\sup_{u\in \mathcal{U}} \dot B_a(x) \ge  -\alpha(B_a(x))$ i.e. $B_a(x)$ is a CBF of set $C_{\hat\theta}^s = \left\{ B_a(x) > 0\right\}$. Therefore, $C_{\hat\theta}^s$ is forward invariant. The proof is done.
\par The CBF based safe adaptation can be obtained with a point-wise solver using quadratic programming (QP). Specifically, the robust safe adaptation QP (RSA-CBF-QP) includes three parts, i.e. the objective function, the safe constraints describe by robust adaptive CBF and control input constraints. The total algorithm including finite-time RSA-CBF-QP is given in Algorithm \ref{algorithm1}.

\begin{algorithm}[t]
	\caption{Finite-time RSA-CBF-QP}
	\label{algorithm1}
	$\textbf{Initialization:}$ estimation $\hat \theta(0)$, worst-case error $\tilde \psi(t)$, signals $\delta = 0$, $\kappa = 0$, filters $\mathcal{H}$ and reference control input $u_{ref}(t)$. 
	\par $\textbf{Parameter setting:}$ learning rate $\gamma$, high gain coefficient $r$ and IE parameter $\beta$ (related to end time $t_c$).
	\par \textbf{Robust safe adaptation:}
	\par $\quad$\textbf{While} $\psi(t) > 0$
	\par $\quad$$\quad$ Sample state and calculate signals $\delta(t)$, $\kappa(t)$ and $\psi(t)$ from systems \eqref{system_dynamics}.
	\par $\quad$$\quad$ Update estimation parameters $\hat \theta$ according to finite-time adaptation law \eqref{finitetime_law}.
	\par $\quad$$\quad$ Modify reference control input $u_{ref}$ based on following RSA-CBF-QP:
	\begin{gather*}
		\text{Objective:}\quad \min \norm{ u - u_{ref} }^2\\
		\text{s.t.}\\ 
		\left\{\begin{aligned}
		\text{Adaptive CBF Safe Constraint: \eqref{th1_acbf}, \eqref{coro1_acbf} or \eqref{thm2_racbf}}\\
		\text{Control Input Constraint:} \quad u_{min} \leq u \leq u_{max}\\
		\text{Other Constraints: e.g. CLF constraints.}
		\end{aligned}\right.
	\end{gather*}
	\par $\quad$$\quad$ Update input $u$ to control systems.
	\par $\quad$\textbf{End While}
	\par $\textbf{Output:}$ estimated parameters $\hat \theta$.
\end{algorithm}

\section{Simulations} \label{section_sim}
\begin{table}[t]
	\caption{Parameter Setting in ACC example.}
	\centering
	\begin{tabular}{|cc|cc|}
		\hline
		$m$ & $1600kg$ & $g$ & $9.81m/s^2$ \\
		\hline
		$v_0$ & $10m/s$ & $f_0 / m$ & $(0.0,~1.962)m/s^2 $\\ 
		\hline
		$f_1$ & $(0.0,~0.002){Ns}/{m}$ & $f_2$ & $(0.0,~0.001){Ns^2}/{m^2}$\\
		\hline
		$z(0)$ & $50m$ & $v(0)$ & $10m/s$\\
		\hline
		$u_{\min}$ & $-0.4mg$ & $u_{\max}$ & $0.4mg$ \\
		\hline
	    $T_h$ & $1.8s$ &$u_{ref}$ & $(0.15 + 0.1sin(t))mg $ \\
		\hline
		$r$ & $0.5$ &$\gamma$ & $2.0 $ \\
		\hline
	\end{tabular}
	\label{table_param_setting_example1}
\end{table}

\begin{figure}[t]
	\centering
	\subfigure[]{
		\begin{minipage}{8.1cm}
		\centering
		\includegraphics[width=8.1cm]{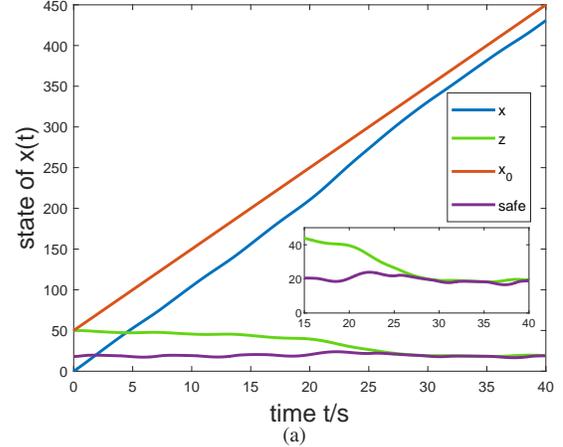}
		\end{minipage}
		\label{figacc_position}
		}
	\subfigure[]{
		\begin{minipage}{8.1cm}
		\centering
		\includegraphics[width=8.1cm]{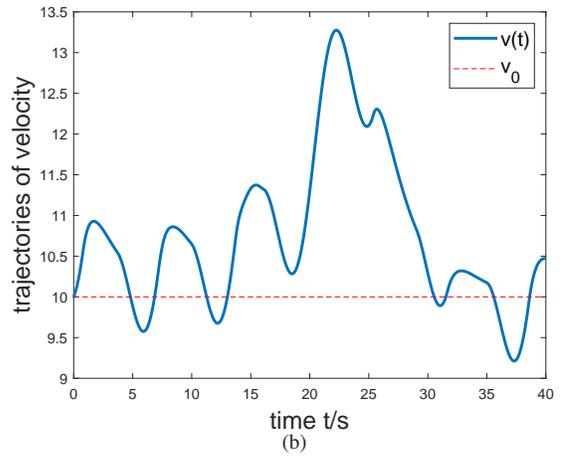}
		\end{minipage}
		\label{figacc_velocity}
		}
		\caption{Positions and velocities of controlled vehicle $x$, $v$ and head vehicle $x_0$, $v_0$. (a) Positions of the two vehicles. The distance between them is $z$, which is varying according to control input. The safe distance, namely $B(x)=0$ is shown by purple line. It can be seen that the safe distance is always small than the real distance, i.e. ACC system is safe. (b) The red dotted line is the velocity of head vehicle $v_0 = 10$, while the velocity of controlled vehicle $v(t)$ is varying according to the control input. When $v(t)>v_0$, the distance $z$ decreases; when $v(t)<v_0$, $z$ increases. The RSA-CBF-QP algorithm always keeps the system in the safe place.}
		\label{figacc_state}
\end{figure}
\begin{figure}[t]
	\centering
	\subfigure[]{
		\begin{minipage}{8.1cm}
		\centering
		\includegraphics[width=8.1cm]{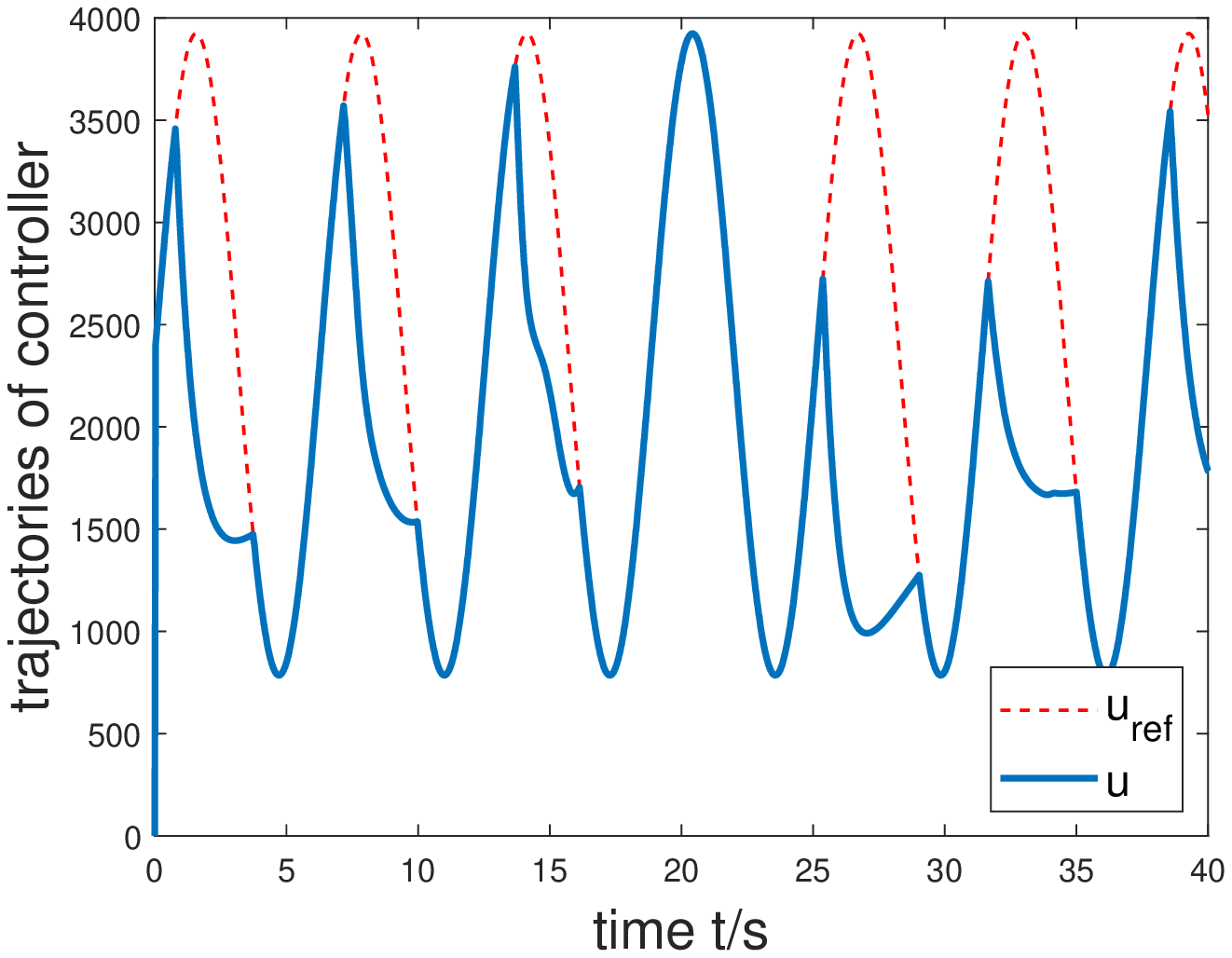}
		\end{minipage}
		\label{figacc_input_ref}
		}
	\subfigure[]{
		\begin{minipage}{8.1cm}
		\centering
		\includegraphics[width=8.1cm]{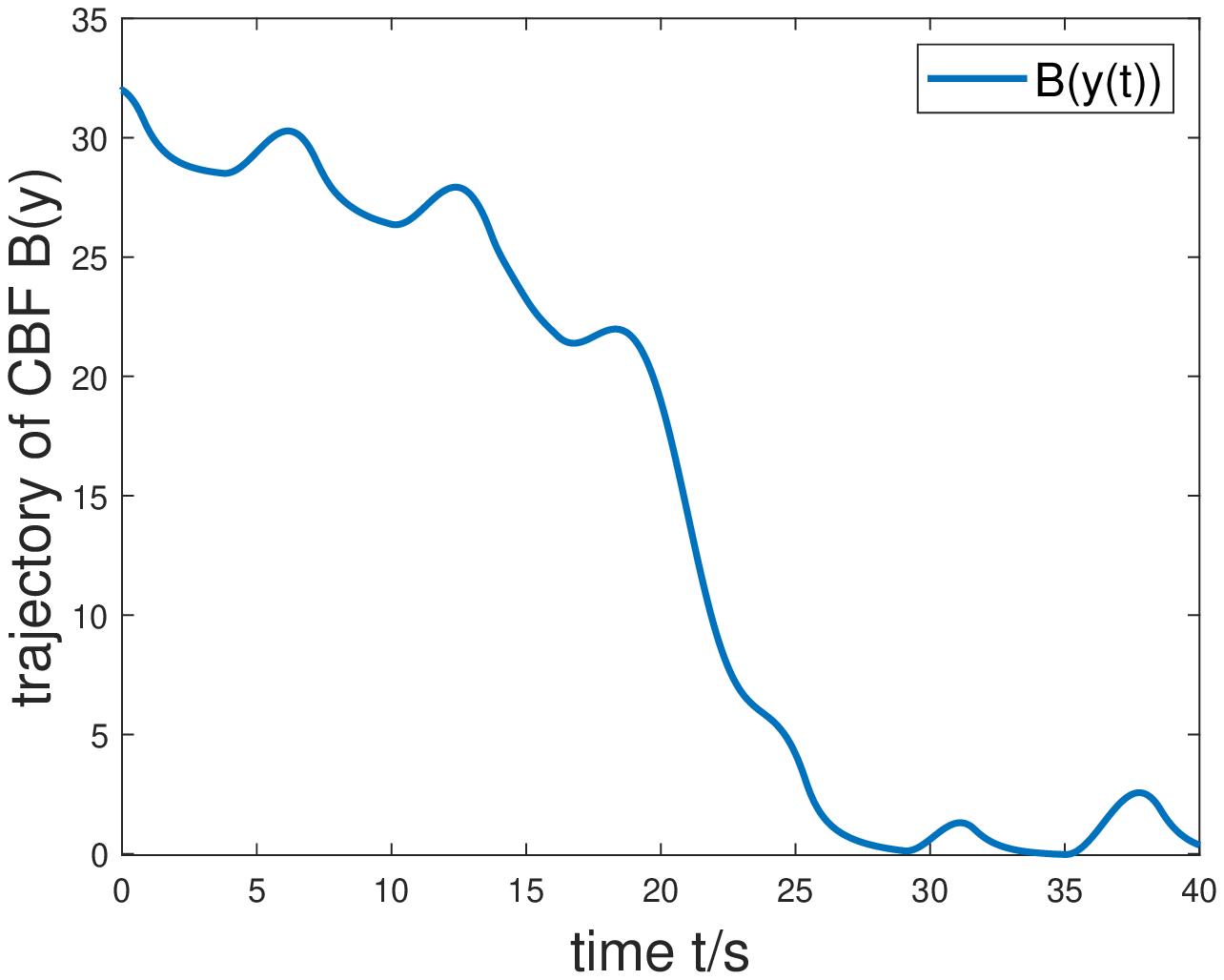}
		\end{minipage}
		\label{figacc_CBF}
		}
		\caption{RSA-CBF-QP algorithm with adaptation input reference. (a) The reference input is generated to ensure the IE condition, through RSA-CBF-QP, the real input is modified to ensure safety. At the beginning, even if the distance is large, input is reduced due to the worst-case estimation error. Afterwards, control input is modified since the small distance is closed to the safe distance. (b) The CBF $B(y)$ should always greater than $0$ so long as $y(0) \in {C}$ where $C$ is the safe set. When $B(y)$ is closed to zero, RSA-CBF-QP enforces $B(y)\ge0$. The trajectory implies the vehicle controlled by noise injected input approaches to the head vehicle during adaptation process, the safe algorithm keeps it away from the safe distance.}
		\label{figacc_control_CBF}
\end{figure}
\begin{figure}[t]
	\centering
	\subfigure[]{
		\begin{minipage}{8.1cm}
		\centering
		\includegraphics[width=8.1cm]{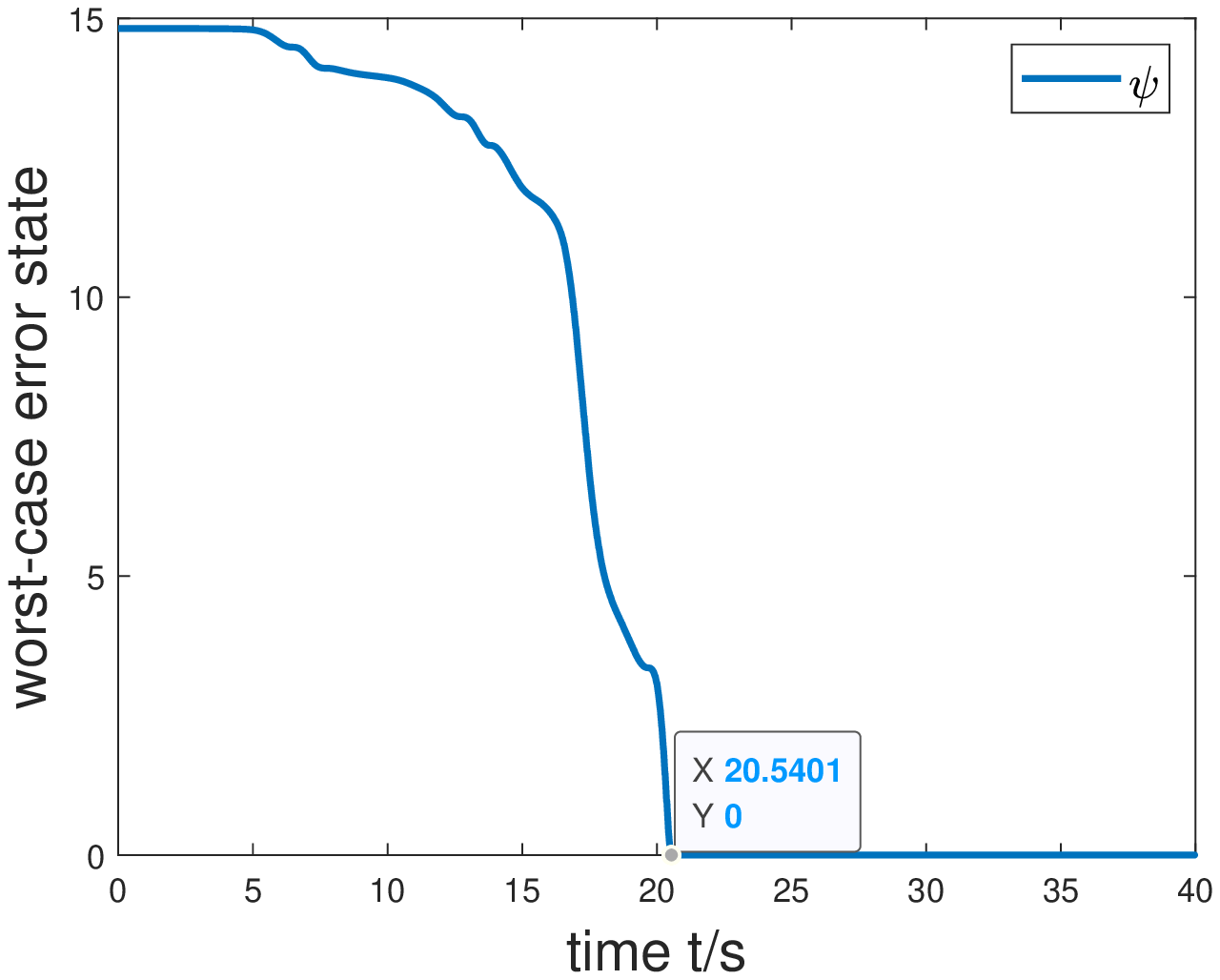}
		\end{minipage}
		\label{figacc_worstcase}
		}
	\subfigure[]{
		\begin{minipage}{8.1cm}
		\centering
		\includegraphics[width=8.1cm]{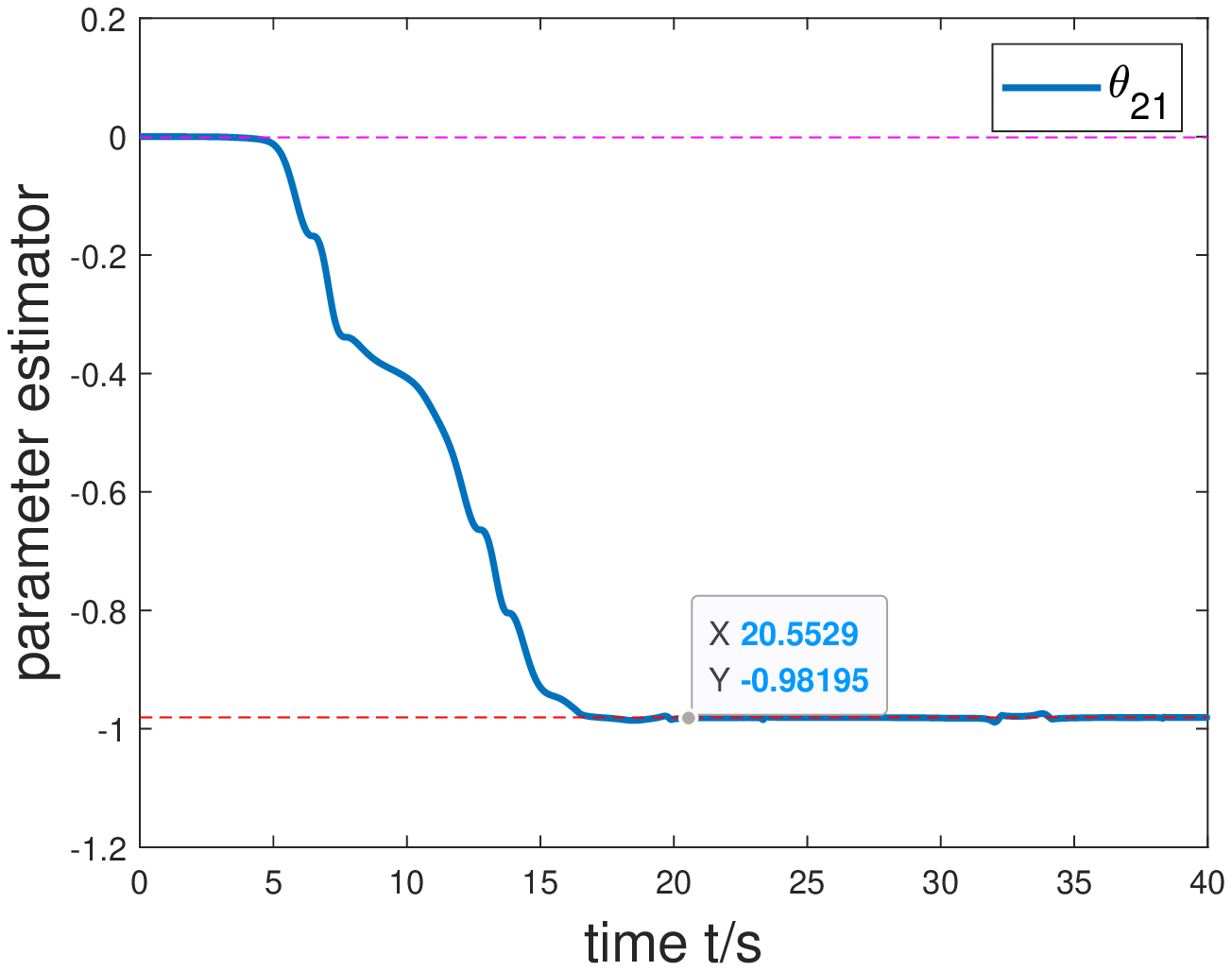}
		\end{minipage}
		\label{figacc_estimator}
		}
		\caption{DREM algorithm for estimation of slope resistance parameter $f_0/m$ in \eqref{acc_dynamics}, considering $f_1$ and $f_2$ to be small disturbances. (a) The worst-case estimation error as \eqref{th1_cbf_adaptive_item}. With the finite-time adaptation law, the worst-case error reduced to $0$ within $21s$ monotonically when satisfying IE condition. (b) The estimation of slope resistance. the red dotted line gives the real value if $f_0/m$, while the pink lines give the unknown value of $f_1$ and $f_2$, which is rather small. It is shown that within $17s$, the parameter can be estimated with acceptable accuracy. The final estimation value of $f_0/m = -0.98195$ since the finite-time algorithm ended at this point. The small perturbation is due to the unknown disturbances, which reflects the robustness of the proposed algorithm.} 
		\label{figacc_DREM}
\end{figure}
\begin{figure}[t]
	\centering
	\subfigure[]{
		\begin{minipage}{8.1cm}
		\centering
		\includegraphics[width=8.1cm]{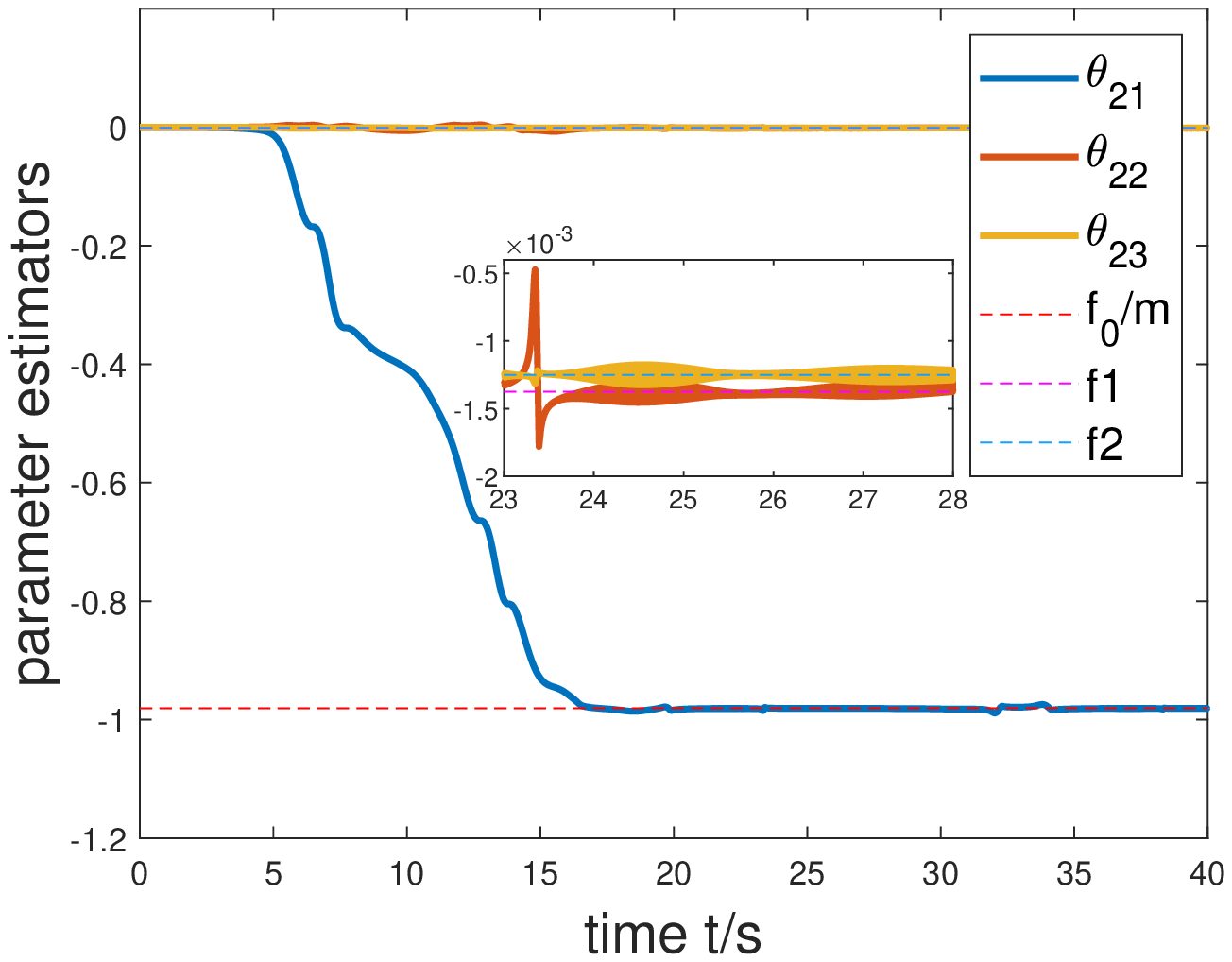}
		\end{minipage}
		\label{figacc_10000hz}
		}
	\subfigure[]{
		\begin{minipage}{8.1cm}
		\centering
		\includegraphics[width=8.1cm]{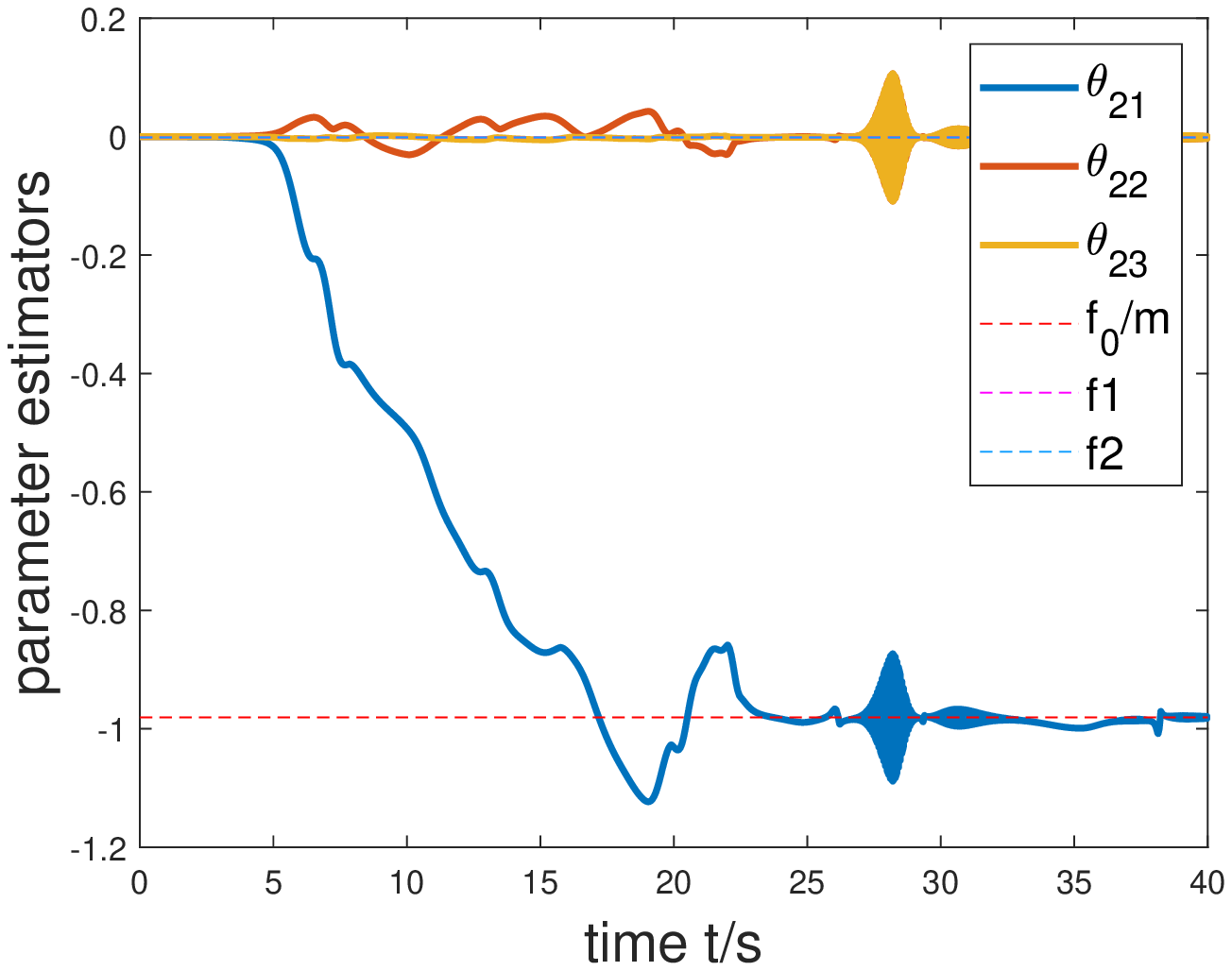}
		\end{minipage}
		\label{figacc_1000hz}
		}
		\caption{DREM algorithm for estimation of all three unknown parameters in \eqref{acc_dynamics}. (a) Estimation with sampling frequency 10000Hz. It is shown that the small parameters $f_1$ and $f_2$ can be estimated with acceptable bounded error. However, perturbation exists due to the error brought from discrete sampling. (b) Estimation with sampling frequency 1000Hz. Although the large parameter $f_0 / m$ is estimated within acceptable error bound, the estimation results of $f_1$ and $f_2$ are unsatisfactory. The monotonicity of DREM dismisses due to the discrete sampling.}
		\label{figacc_compare}
\end{figure}

In this section, a practical problem, namely adaptive cruise control (ACC) that has been studied in \cite{ACC_example} is considered to show the effectiveness of proposed algorithms. The modeling process is followed but the model parameters change from known to unknown. One different setting from \cite{ACC_example} is that the slope resistance related to the gravity of vehicle is considered thus it can no longer be treated as disturbance and needs to be estimated. Specifically, consider a vehicle moving along a straight line (with an unknown slope) with kinematic dynamics
\begin{equation}
	\left[\begin{matrix}
		\dot v\\
		\dot x
	\end{matrix}\right] = \left[\begin{matrix}
		-\frac{1}{m}F_r(v) + F_u\\
		v
	\end{matrix}\right], \label{acc_vehicle_dynamics}
\end{equation}
where $v$ and $x$ represent the velocity and position of the vehicle, respectively. The mass of the vehicle is assumed to be known and  notated as $m$. $F_u = \frac{1}{m}u$ with $u$ be the control input of wheel force. $F_r(v) = f_0 + f_1v + f_2 v^2$ represent the resistances with slope resistance $f_0$ and rolling resistance $f_1$ and $f_2$, which were empirically set such that the dynamics \eqref{acc_vehicle_dynamics} is totally known in \cite{ACC_example}. Differently, the coefficients of $F_r(v)$ are assumed to be unknown, only with given possible ranges. Notice that the slope resistance is related to the mass of vehicle, it may have a great influence to control performance and should be estimated, while $f_1$ and $f_2$ are rather small which can be treated as disturbances. Another vehicle in front of the controlled one \eqref{acc_vehicle_dynamics} is moving with a constant velocity $v_0$. The distance between two vehicles is denoted by $z$. Therefore, the dynamics of ACC problem can be formulated as
\begin{equation}
	\dot y = \underbrace{\left[\begin{matrix}
		v \\
		0\\
		v_0 - v
	\end{matrix}\right]}_{f(y)} - \underbrace{\left[\begin{matrix}
		0& 0& 0 \\
		f_0 / m& f_1& f_2\\
		0& 0&   0
	\end{matrix}\right]}_{\theta}\underbrace{\left[\begin{matrix}
		1 \\
		v\\
		v^2
	\end{matrix}\right]}_{\Delta(y,u)} + \underbrace{\left[\begin{matrix}
		0 \\
		\frac{1}{m}\\
		0
	\end{matrix}\right]}_{g(y)} u. \label{acc_dynamics}
\end{equation}
where $y = [x, v, z]^\Tran$. The control objective is trying to track a reference control input $u_{ref}$ (to excite systems for adaptation) without obeying the hard constraints: (1) the distance between two vehicles must be greater than the safe distance $D_s(v)$, e.g. $D_s(v) = T_h v$ in which $T_h$ represents the look ahead time; (2) the control input, namely the acceleration is limited as $u_{\min} \leq u \leq u_{\max}$. Let $g$ be the acceleration of gravity.
\par To convert the safety constraints, a CBF can be designed as $B(y) = z - T_h v$ which implies $C=\left\{B(y)\ge0\right\}$ is a safe set. However, this may conflict with the input constraints, as discussed in \cite{ACC_example}. Therefore, a modified CBF considering the input limitations can be constructed as
\begin{equation}
	B(y) = z - T_hv - \frac{1}{2u_{\min}}(v - v_0)^2.
\end{equation}
Also, the safe set is $C=\left\{B(y)\ge0\right\}$. To ensure the signal $\delta(t) \notin \mathcal{L}_2$ or satisfying IE condition, noises are always added to control input, randomly or elaborately. In this simple example, a time-dependent sinusoidal generator is injected to the desired control input, which is denoted as $u_{ref}$. Detailed parameter setting can be found in Table \ref{table_param_setting_example1}. In simulations, the true values of unknown parameters is $f_0/m = 0.981$, $f_1 = 0.0013$ and $f_2 = 0.00125$.
\par To keep system safe, a RSA-CBF-QP algorithm is designed as shown in Algorithm \ref{algorithm1}, in which the objective function is $\min \norm{u - u_{ref}}^2$ and only $f_0 /m $ is estimated, $f_1$ and $f_2$ are treated as bounded, unknown disturbances. The real control input is computed with frequency $100Hz$, which can be satisfied using QP toolbox. The initial value of estimator is $0$. The positions, velocities and distance in safe adaptation process can be seen in Fig \ref{figacc_state}, which implies that the safety is always ensured. To show the details on how RSA-CBF-QP works, Fig \ref{figacc_control_CBF} shows the modification of control input and the real time value of CBF $B(y)$. The RSA-CBF-QP promises that if $y(0) \in C$, then $s(t) \in C$ for all $t>0$. The key process of safe adaptation is parameter estimation, with finite-time DREM, the estimation process is given in Fig \ref{figacc_DREM}. It is shown that the unknown parameter $f_0/m$ is estimated successfully within finite time (the maximum estimation time is determined by the worst-case error shown in Fig \ref{figacc_worstcase}). The sampling frequency is $10000Hz$ to approximated the continuous regression process as DREM needs. The slight perturbations of estimation is due to the unknown disturbances. 
\par As a more practical problem, the slope resistance is considered in this example and a robust safe adaptation algorithm is used to estimate the related parameter. The conservativeness is mainly reflected in two aspects. On the one hand, the worst-case estimation error is considered, resulting in a discount in control objective, e.g. $\min \norm{u - u_{ref}}^2$. It is shown in Fig \ref{figacc_control_CBF} that even the $B(y)$ is far from $0$, $u(t)$ can take the value of $u_{ref}$ without violation of safe constraints, i.e. $B(y) \leq 0$ (as shown in the time interval $(17s, 23s)$ in Fig \ref{figacc_input_ref}). This is due to the predefined range is rather broad. If so, the performance will be discounted greatly at the beginning of adaptation. This impact can be reduced by giving more accurate bounds. A more complex situation of cruise control was studied in \cite{Safer_Navigation_TCYB}, where the conservativeness from only considering the kinematics was extended by calculating the time to collision (TTC). Combining efficient CBF with TTC can be a valuable further topic. On the other hand, the process of (finite-time) DREM needs high frequency sampling of system state (whereas the CBF-based quadratic programming can be computed within $100Hz$). If the frequency is low, the effect of estimation will deteriorate, especially when a high-gain finite-time law is utilized as \eqref{finitetime_law}. To verify this, a comparison is made using different sampling frequency of $1000Hz$ and $10000Hz$. In this comparison experiment, all unknown parameters need to be estimated to eliminate the influence of disturbances. As a result, the estimation effects are illustrated in Fig \ref{figacc_compare}. Therefore, although the robustness can be describe quantitatively using high-gain adaptive law, it is not recommended to directly use it with a low sampling frequency. Alternatively, finite-time DREM without introducing high gains \cite{matrix_estimation_finite_time} can also be considered, while its robustness still needs further research to highlight its advances compared to high-gain adaptation \cite{finite_time_robust}.

\section{Conclusion}
This paper has studied the safety-guaranteed parameter identification algorithms by combination finite-time DREM and robust aCBF method. The control input intended to active the control systems is filtered by the aCBF-based quadratic programming to keep the system in a predefined safe region. The robustness of this algorithms is analyzed, and the conservativeness of the safety constraints is further relaxed. A simplified ACC example with unknown parameters to be determined is simulated to show the effectiveness of the presented methods. To further improve the algorithms, the adaptive CLF \cite{robust_cbf,fixedtimeACBF,safe_uncertain_ELsystem} can be designed to achieve certain control objectives, of which the theoretical framework needs to be analyzed for feasibility and stability, as studied in \cite{CLFCBF_extended}. Besides, the safe identification of time-varying parameters may be another interesting research with more relaxed safety constraints.

{	\footnotesize
\bibliographystyle{IEEEtran}
\bibliography{mybib}
}

\end{document}